\documentclass[aps,prl,reprint,superscriptaddress]{revtex4-1}

\usepackage{amsmath,amssymb,graphicx}
\usepackage{xcolor}
\usepackage[colorlinks=true,urlcolor=blue,linkcolor=blue,citecolor=blue,bookmarks=false]{hyperref}

\newcommand{\frmlDot}{\;\text{.}}
\newcommand{\frmlComma}{\;\text{,}}
\newcommand{\figwidth}{1}

\begin{document}

\title{Spin structure relation to phase contrast imaging of isolated magnetic Bloch and N\'eel skyrmions}

\author{S. P\"ollath}
\affiliation{Institut f\"ur Experimentelle Physik, Universit\"at Regensburg, D-93040 Regensburg, Germany}

\author{T. Lin}
\affiliation{Fert Beijing Institute, BDBC, School of Microelectronics, Beihang University, Beijing, 100191, China}

\author{N. Lei}
\affiliation{Fert Beijing Institute, BDBC, School of Microelectronics, Beihang University, Beijing, 100191, China}

\author{W. Zhao}
\affiliation{Fert Beijing Institute, BDBC, School of Microelectronics, Beihang University, Beijing, 100191, China}

\author{J. Zweck}
\affiliation{Institut f\"ur Experimentelle Physik, Universit\"at Regensburg, D-93040 Regensburg, Germany}

\author{C.H. Back}
\email[e-mail:]{christian.back@tum.de}
\affiliation{Physik-Department, Technische Universit\"at M\"unchen, D-85748 Garching, Germany}
\affiliation{Munich Center for Quantum Science and Technology (MCQST), Schellingstr. 4, D-80799 M\"unchen}


\date{\today}

\begin{abstract}
Magnetic skyrmions are promising candidates for future storage devices with a large data density. A great variety of materials have been found that host skyrmions up to the room-temperature regime. Lorentz microscopy, usually performed in a transmission electron microscope (TEM), is one of the most important tools for characterizing skyrmion samples in real space. Using numerical calculations, this work relates the phase contrast in a TEM to the actual magnetization profile of an isolated N\'eel or Bloch skyrmion, the two most common skyrmion types. Within the framework of the used skyrmion model, the results are independent of skyrmion size and wall width and scale with sample thickness for purely magnetic specimens. Simple rules are provided to extract the actual skyrmion configuration of pure Bloch or N\'eel skyrmions without the need of simulations. Furthermore, first differential phase contrast (DPC) measurements on N\'eel skyrmions that meet experimental expectations are presented and showcase the described principles. The work is relevant for material sciences where it enables the engineering of skyrmion profiles via convenient characterization.
\end{abstract}

\maketitle

\section{Introduction}

Magnetic skyrmions are tiny magnetic spin whirls that can be found in materials with a crystal structure that exhibits a broken inversion symmetry and spin orbit interaction. These two ingredients enable antisymmetric exchange interaction between neighboring spins which leads to the unique spin structure of the skyrmion \cite{Dzyaloshinsky1958,Moriya1960,2009_Muhlbauer_Science}. This particular topologically protected spin structure is responsible for the skyrmion's strong coupling to external stimuli like magnetic fields, electric or heat currents \cite{White2014, 2016_Ehlers_PhysRevB,Fert2013,Poellath2017}. Additionally, the structure awards the skyrmion with a robustness due to its inherent topological protection \cite{2013_Milde_Science, 2017_Makino_PhysRevB, 2017_Wild_SciAdv}. 

All these properties make the skyrmion a potential candidate for future storage devices. Their size in the nm-range in combination with their topological protection in principle allows a large information density. Substantial scientific efforts were made to find new skyrmion-hosting materials with engineered properties \cite{2010_Munzer_PhysRevB,Yu2011,Soumyanarayanan2017, 2017_Karube_PhysRevMater, 2017_Nayak_Nature,McVitie2018}. One of the most important tools for real-space characterization of skyrmions is Lorentz microscopy that is usually performed in a Transmission Electron Microscope (TEM). Due to its high spatial resolution and sensitivity to magnetic fields, the technique is well suited for metrology of magnetic skyrmions. 

One aspect of TEM measurements on skyrmions that has to be considered carefully is that the contrast formation is rather complex as the electron beam completely passes through the thin specimen and is also affected by its magnetostatic (stray-)fields. This means that the obtained TEM images are not directly interpretable when the actual skyrmion spin structure is of interest. This work directly addresses this issue and relates the electron phase contrast from TEM to the magnetization profiles of isolated Bloch and N\'eel skyrmions, which are the two most common skyrmion types. It is shown that the results are applicable to any skyrmion size, saturation magnetization and are expected to hold for samples that have a thickness in the order of the electron's mean free path. Besides the detailed results from the calculations, convenient rules are presented that enable the experimentalist to quickly determine the skyrmion radius and wall width. These results can support the engineering and optimization of skyrmion materials as they enable a fast and reliable characterization. Furthermore, many theoretical estimations of skyrmion properties like skyrmion lifetime, topological Hall-angle or magnetic resonance frequencies rely on precise measurements of the skyrmion structure \cite{Bessarab2018,Sampaio2013, Buettner2018, Schwarze2015, Mochizuki2012}. Also, the ratio of skyrmion radius to wall width can be a strong indicator for the distinction between a so-called circular chiral bubble domain and a skyrmion. Finally, first Differential Phase Contrast (DPC) measurements on N\'eel skyrmions, that meet theoretical expectations, are presented and are used to showcase the application of the theoretical results.

\section{Calculated electron phase for Bloch skyrmions}

The skyrmion model of Büttner et al.\ is used to generate the skyrmion structure \cite{Buettner2018}. The model expresses the normalized magnetization vector field $\mathbf{m}(\mathbf{r})=\mathbf{M}(\mathbf{r})/M_s$ that describes a radially symmetric isolated skyrmion with the cylindrical coordinates $r\text{,}\varphi$ as


\begin{align}
\mathbf{m}(r,\varphi)=&\sin (\theta) \cos (\varphi+\psi) \hat{\mathbf{e}}_{x}+\sin (\theta) \sin (\varphi+\psi) \hat{\mathbf{e}}_{y}\nonumber\\
&+\cos (\theta) \hat{\mathbf{e}}_{z} \label{frml_skstructure1} \frmlComma \\
 \theta\left(r, R^*, \Delta^*\right)=&\;\theta_{\mathrm{DW}}\left(r-R^*, \Delta^*\right)\nonumber\\
 &+\theta_{\mathrm{DW}}\left(r+R^*, \Delta^*\right)-(N+1) \pi / 2 \label{frml_skstructure2} \frmlComma \\
 \theta_{\mathrm{DW}}(r, \Delta^*)=&\;2 \arctan \left(\exp \left(r/\Delta^*\right)\right) \label{frml_skstructure3} \frmlComma \\
 R^* =&\; \Delta^* \ln \left(\sinh(R/\Delta^*)+\sqrt{\sinh ^{2}(R/\Delta^*)-1}\right) \label{frml_skstructure4} \frmlComma \\
 \Delta^* =&\; 0.12 \cdot \Delta \label{frml_skstructure5}\frmlDot
 \end{align}

Here, $R$ is the skyrmion radius with $m_z(R)=0$; $\Delta$ is the experimentally relevant domain wall width defined by the radial distance from $r=R_i$ to $r=R_o$ with $m_z(R_i)=N\cdot 0.99$ and $m_z(R_0)=-N\cdot 0.99$ (numerically evaluated). The skyrmion polarity $N=\pm 1$ defines the sign of $m_z$ in the skyrmion core. The helicity $\psi$ determines the skyrmion type which is $\psi=0,\pi$ for N\'eel and $\psi=\pi/2\text{,}\;3\pi/2$ for Bloch skyrmions of the two respective winding possibilities.

\begin{figure}
	\includegraphics[width=\figwidth\linewidth]{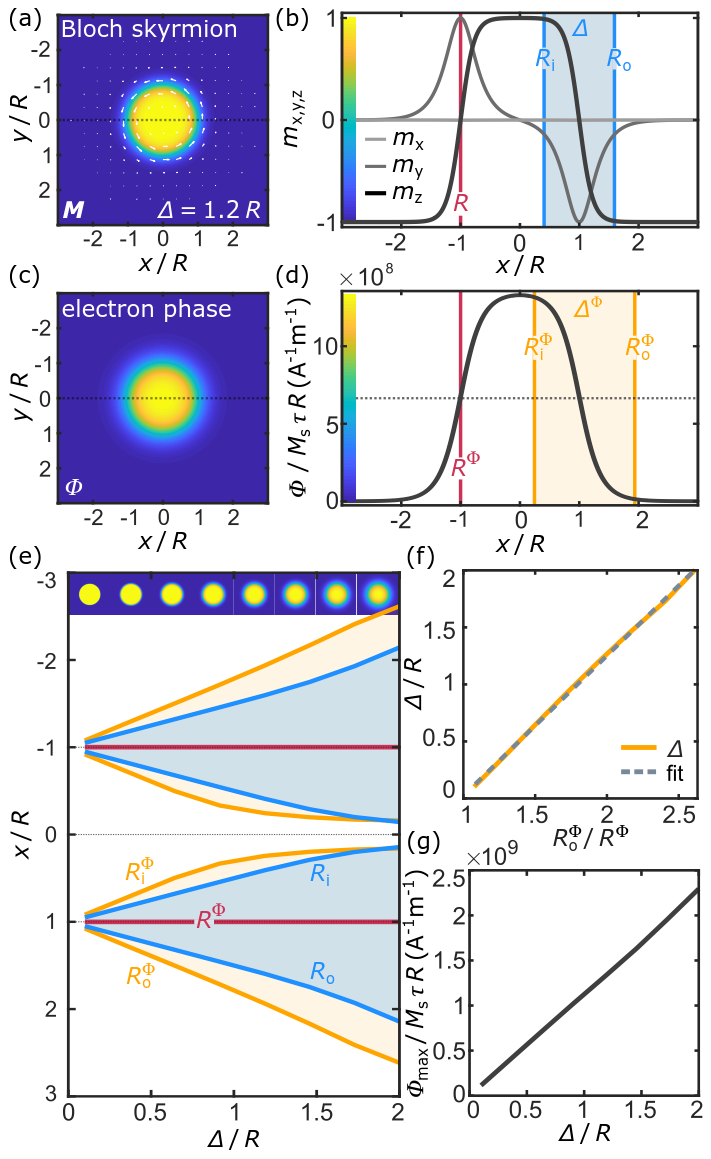}	
	\caption{\label{figure1}Relation of the Bloch skyrmion structure to the electron phase. 
		\textbf{(a)} Example Bloch skyrmion structure. Color represents the $m_z$-component as indicated in (b).
		\textbf{(b)} Line profile of the magnetization components along the dotted line in (a).
		\textbf{(c)} Calculated electron phase of (a) under normal electron incidence. Color represents the phase value as indicated in (d).
		\textbf{(d)} Line profile of the electron phase along the dotted line in (c).
		\textbf{(e)} Dependence of $R_i$, $R_o$, $R_i^\Phi$, $R_o^\Phi$ and $R^\Phi$ on the domain wall width $\Delta$.
		\textbf{(f)} Dependence of $\Delta$ on $R_o^\Phi$ and linear fit.
		\textbf{(g)} Dependence of the total phase change $\Phi_\text{max}$ introduced by the skyrmion with respect to domain wall width.
	}
\end{figure}

Fig.~\ref{figure1}(a) shows the exemplary 2D spin structure for an isolated Bloch skyrmion with $\psi=0$, $N=1$ and $\Delta=1.2 R$. Color represents the $m_z$ component as indicated at the y-axis of Fig.~\ref{figure1}(b). The line profile of the magnetization components $m_{x,y,z}$ along the dotted line is shown in Fig.~\ref{figure1}(b) and shows the structural parameters described before. When an electron beam passes through a thin magnetic sample structure, it picks up a spatially varying phase. To calculate this phase modulation $\Phi(x,y)$ for a magnetic skyrmion, a well established method by M.\ Mansuripur is used \cite{Mansuripur2019}. For a given magnetization and with the assumption of a thin sample, the method basically calculates the vector potential and also accounts for stray fields outside of the sample. We will exploit that Mansuripur's result can be rewritten in a way that $\Phi$ can be normalized by the saturation magnetization $M_s$, the sample thickness $\tau$ and the lateral sample dimensions or in this case the skyrmion radius $R$.

The resulting electron phase $\Phi$ for the skyrmion structure of Fig.~\ref{figure1}(a) is shown in Fig.~\ref{figure1}(c). The coloring represents normalized phase values as again indicated in the line profile shown in Fig.~\ref{figure1}(d). Although $m_z$ does not contribute to the electron phase contrast, $\Phi$ shows striking similarities to $m_z$. For the given domain wall width, the phase's point of inflection $R^\Phi$ (indicated by the red line) coincides with the skyrmion radius. The radius $R_i^\Phi$ where the phase is $1\%$ of its maximum and the radius $R_o^\Phi$ where the phase reaches $99\%$ of its maximum value (indicated by orange lines) have a larger radial distance $\Delta^\Phi$ to $R^\Phi$ than their spin structural counterparts $R_i$ and $R_o$ shown in Fig.~\ref{figure1}(b). 

Latter relations are now discussed for all physical domain wall widths, i.e.\ from values near zero, corresponding to sharp domain walls and a magnetic bubble like structure, up to domain wall widths of twice the skyrmion radius where the two radially opposing domain walls start to overlap. The results are summarized in Fig.~\ref{figure1}(e). The narrow horizontal image strip at the top of the plot indicates the $m_z$ configuration of the skyrmion for the respective wall width $\Delta$ on the x-axis. The red colored lines show the position of $R^\Phi$ with respect to $\Delta$. It can be seen that $R^\Phi$ coincides with $R$ for all domain wall widths. This means that the Bloch skyrmion radius can directly be determined from the phase contrast measurement by measuring the point of inflection of the phase. Alternatively, if the point of inflection is experimentally not well accessible, the radius where the phase reaches half of its maximum value can be measured which will only lead to a maximum overestimation of the skyrmion radius by $7\%$ for largest domain wall widths (graph not shown). The blue and orange shaded areas relate the domain wall parameters $R_i$, $R_o$ and $\Delta$ with the phase structure parameters $R_i^\Phi$, $R_o^\Phi$ and $\Delta^\Phi$ as indicated in Fig.\ref{figure1}(b) and (d). As seen before, the phase domain wall width $\Delta^\Phi$ extends further out than $\Delta$. Unfortunately, there is no linear relation between $\Delta^\Phi$ and $\Delta$. However, it turns out that the dependence of $\Delta$ on $R_o^\Phi$ is linear for the whole range as indicated by the linear fit in Fig.~\ref{figure1}(f). Finally, Fig.~\ref{figure1}(g) shows the normalized phase value from its base $\Psi(r\rightarrow\infty)$ to maximum level $\Psi(0)=\Psi_\text{max}$ with respect to the domain wall width which can be used for quantitative phase measurements. As expected, the total phase gain increases with domain wall width as the phase is generated from the in-plane magnetic wall.

\begin{figure} [tb]
	\includegraphics[width=\figwidth\linewidth]{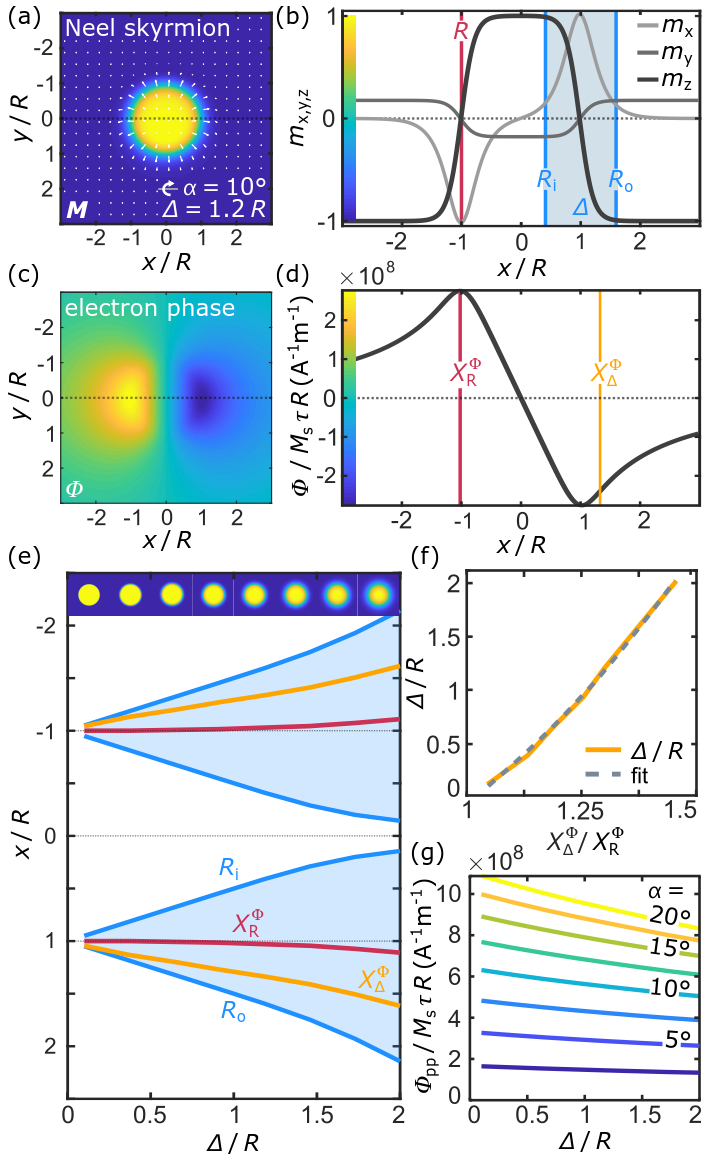}	
	\caption{\label{figure2}Relation of the N\'eel skyrmion structure to the electron phase. 
		\textbf{(a)} Example N\'eel skyrmion structure. Color represents the $m_z$-component as indicated in (b).
		\textbf{(b)} Line Profile of the three magnetization components along the dotted line in (a).
		\textbf{(c)} Calculated electron phase of (a) under $10^\circ$ electron incidence. Color represents phase values as indicated in (d).
		\textbf{(d)} Line profile of the electron phase along the dotted line in (c).
		\textbf{(e)} Dependence of $R_i$, $R_o$, $X_R^\Phi$ and $X_\Delta^\Phi$ with respect to the domain wall width $\Delta$.
		\textbf{(f)} Dependence of $\Delta/R$ with respect to $X_\Delta^\Phi / X_R^\Phi$ and quadratic fit.
		\textbf{(g)} Peak to peak phase difference $\Phi_{pp}$ in dependence of $\Delta$ for different tilt angles $\alpha$.
	}
\end{figure}

In summary, in a TEM experiment that retrieves the electron phase, the Bloch skyrmion radius $R$ can be obtained from the point of inflection of the phase, or approximately from the radius where the phase reaches half of its maximum, i.e. $R = R^\Phi$. The skyrmion domain wall width $\Delta$ is then obtained from any relation in Fig.~\ref{figure1}(e) or conveniently from the linear relation $\Delta / R = 1.23 \cdot R_o^\Phi / R^\Phi -1.21$. 

Naturally, the experimental features of the phase are not limited to bright field imaging techniques like Electron Holography or Fresnel Lorentz TEM, but can also be transferred to Differential Phase Contrast techniques performed in the Scanning mode of the TEM (STEM). An electron probe size $R_\text{probe} = 10\%\cdot R$, will only lead to an additional error of Bloch skyrmion radius estimation of $1\%$ and an absolute overestimation of $\Delta/R$ by $0.25$ but only for smallest $\Delta$ in a center of mass measurement, as additional simulations have shown. 

Further note, that the results apply to the recently proposed skyrmion surface states which should make up significant parts in thin TEM samples \cite{Zhang2018, Legrand2018}. The spin configuration of such a surface state is expressed by equations \ref{frml_skstructure1}-\ref{frml_skstructure5} and a $z$-dependent skyrmion helicity $\psi(z)$ that deviates from the values discussed for Bloch and N\'eel skyrmions at the sample surface. It can be shown that in this case, the spin structure can be written as a superposition of Bloch and N\'eel skyrmion. As N\'eel skyrmions at zero sample tilt do not contribute to the electron phase, as described later, the skyrmion edge state will therefore only lead to the introduction of a reduced effective thickness $\tau^* = \int_{-\tau/2}^{\tau/2} \left|\sin(\psi(z))\right|dz$.

\section{Calculated electron phase for N\'eel skyrmions}

A very similar discussion can be made in the case of the N\'eel skyrmion. There is however, one issue that has to be addressed in the beginning. As the z-component of the curl of the N\'eel skyrmion's magnetization vanishes, traversing electrons do not get phase modulated. This is why N\'eel skyrmions cannot be detected by electron phase contrast microscopy for normal beam incidence \cite{McVitie2003}. Therefore it is required to tilt the sample with respect to the incoming electron beam which breaks the rotational symmetry of the problem. Fig.~\ref{figure2} shows the 2D skyrmion structure of a N\'eel skyrmion with $\Delta = 1.2 R$, $N=1$ and $\psi = 3\pi/2$ that was tilted by $\alpha=10^\circ$ around the horizontal axis as indicated. As before, the dashed line indicates the line profiles of the magnetization shown in Fig.~\ref{figure2}(b).

The respective electron phase is shown in Fig.~\ref{figure2}(c) and features an area of positive and negative phase along the tilt axis. In contrast to the Bloch skyrmion, the obtained phase looks very different from  $m_z$, because it is mostly generated from the small in-plane $m_y$-components introduced by the sample tilt. Due to the constant in-plane magnetization outside of the skyrmion, a  phase ramp is present in the calculation. This ramp was manually removed for improved visibility which also needs to be done for corresponding experimental data to compare it to these results. A phase line profile along the tilt axis as indicated by the dotted line in Fig.~\ref{figure2}(c) is shown in Fig.~\ref{figure2}(d). As can be seen, the x-coordinate of maximum or minimum phase $X_R^\Phi$ coincides with $R$. The phase's point of inflection $X_\Delta^\Phi$ can be associated with the skyrmion domain wall width $\Delta$ as shown in the following.

Fig.~\ref{figure2}(e) shows $R_o$, $R_i$, $X_R^\Phi$ and $X_\Delta^\Phi$ in dependence of the domain wall width $\Delta$. As can be seen $X_R^\Phi$ coincides with $R$ rather well with a maximum error of $6.5\%$ for the full range of $\Delta$. The point of inflection $X_\Delta^\Phi$ can be quadratically approximated by $\Delta/R= 3.07 \left(\frac{X_\Delta^\Phi}{X_R^\Phi}\right)^2 - 2.93 \frac{X_\Delta^\Phi}{X_R^\Phi} -0.2$ as shown in Fig.~\ref{figure2}(f). The peak to peak phase change $\Phi_{pp}$ with respect to the domain wall width is shown in Fig.~\ref{figure2}(g), which can again be used for quantitative phase measurements. These results are independent of tilt angles $\alpha < 20^\circ$. It should further be noted that the calculations shown in this work assume a purely magnetic specimen and caution is advised when applying these results to systems or techniques where the electron phase is further modified by e.g.\ multiple scattering, electrostatic fields or strong aberrations.

\section{Experimental application to N\'eel skyrmions}

In the last part of this work, first DPC-measurements on magnetic N\'eel skyrmions that match theoretical expectations are presented. The theoretical results on N\'eel skyrmions discussed before can be directly applied to our measurements which also make them an excellent showcase. For phase contrast measurements on magnetic Bloch skyrmions, already a large body of work exists in terms of electron holography and DPC \cite{McGrouther2016, Schneider2018,Park2014,Shibata2017}. In a DPC-measurement, the electron beam of the TEM is focused onto the sample and scanned across the imaged sample area. The diverging electron beam is deflected for example due to magnetic fields. This 2D deflection is measured and is proportional to the gradient of the electron phase shown before in Fig.~\ref{figure2}(c) \cite{Zweck2016}.

\begin{figure} [tb]
	\includegraphics[width=\figwidth\linewidth]{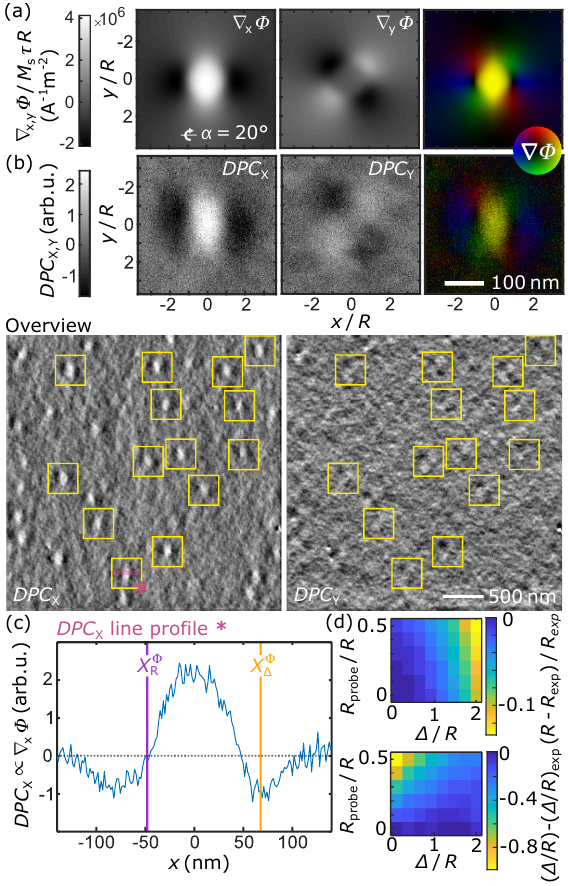}	
	\caption{\label{figure3}Experimental DPC measurement on N\'eel skyrmions compared to theoretical findings. 
		\textbf{(a)} Theoretical 2D maps of the electron phase gradient's components and color encoded phase gradient. Color represents the direction and intensity gradient magnitude.
		\textbf{(b)} Experimental DPC measurements on a multilayer stack of Ta(3)/Pt(5)/[Co(2)/Tb(1)/Pt(3)](5). The shown intensity is the average of 14 skyrmions. On the bottom, a larger field of view is shown and the respective skyrmions used for averaging are marked. For display, these overview images are Gaussian filtered with $\sigma = 2~\text{px}$. The DPC signal is proportional to the electron phase gradient's components.
		\textbf{(c)} Line profile of $\text{DPC}_X$ at $y=0$. The maximum of the phase $X_R^\Phi$ and the point of inflection $X_\Delta^\Phi$ are highlighted. \textbf{D} Correction terms for different electron probe sizes.
	}
\end{figure}

Fig.~\ref{figure3}(a) shows 2D maps of the two components of the normalized phase gradient $\mathbf{\nabla}\Phi$ for a N\'eel skyrmion that is tilted by $20^\circ$ around the horizontal axis and has a polarity of $N=-1$ (compare with Fig.~\ref{figure2}(c)). The gradient's x-component $\nabla_x\Phi$ shows a triple structure along the tilt axis. The deviation from the phase base level is 2.4 times as strong as in the gradient's y-component, that consists of two diagonally opposing pairs of patches with either in- or decreased phase. On the right side of Fig.~\ref{figure3}(a), the phase gradient is shown color-coded, where color represents direction and intensity magnitude of $\nabla \Phi$ as indicated by the color-wheel below. 

For the experiment, a multilayer stack of Ta$_{(3\text{nm})}$/Pt$_{(5)}$[Co$_{(2)}$/Tb$_{(1)}$/Pt$_{(3)}$]$_{\times5}$ is evaporated onto a $500\times 500~\mu m^2$ Si$_3$N$_4$ membrane with a thickness of $\tau_\text{SiN}=20~\text{nm}$. The polycrystalline material has an average grain size of $4.9~\text{nm}$ with a standard deviation of $1.0~\text{nm}$. The specimen is inserted into the TEM of the type FEI Tecnai F30, operated in scanning mode using an electron energy of $300~\text{keV}$. The sample is tilted to $\alpha=20^\circ$ and an external magnetic field of $\mu_0 H = 120~\text{mT}$ is applied along the beam direction. The experiment is conducted at room temperature. The electron beam's deflection is detected using a segmented annular detector. The difference signals of two opposing ring segments are called $\text{DPC}_X$ and $\text{DPC}_Y$. They are proportional to the beam deflection in x- and y-direction and therefore to $\nabla_x\Phi$ and $\nabla_y\Phi$ \cite{Zweck2016, Schwarzhuber2017}. The DPC measurement is performed in a field of view of 11.1 $\mu \text{m}^2$ containing around 30 skyrmions. Due to the crystalline background, the weak phase modulation of the skyrmion is subject to a rather long-period modulated background, as the resolution is above the crystal size. This is why the center of 14 skyrmions is manually selected and an average DPC signal is calculated. The results for $\text{DPC}_\text{X}$ and $\text{DPC}_\text{Y}$ are shown in Fig.~\ref{figure3}(b) and agree well with the theoretical expectations shown above in Fig.~\ref{figure3}(a). Also the color encoded image matches all expected theoretical features. The overview images are shown on the bottom of Fig.~\ref{figure3}(b) and the skyrmions that were used for averaging are marked respectively. Note that in stacked ferromagnet and heavy metal multilayers (like the ones shown here), hybrid skyrmions that are a mixture of Bloch and N\'eel type have been observed \cite{Legrand2018}. As the theory cannot be applied to such tilted structures, it might be a good advice to observe e.g.\ the Fresnel contrast at zero sample tilt which vanishes for a pure N\'eel skyrmion \cite{Fallon2019}. This fact was confirmed for this sample.

A horizontal line profile with an integration width of $65~\text{nm}$ along $\text{DPC}_\text{X}$ of the skyrmion that is marked with a star is shown in Fig.~\ref{figure3}(c). From this profile, the experimental parameters $X_R^\Phi$ and $X_\Delta^\Phi$ can be extracted directly. As $X_R^\Phi$ is the phase maximum, we find it in the $\text{DPC}_\text{X}$ line profile's zero crossing and obtain an experimental skyrmion radius $R_\text{exp}$ of $R=48\pm 3$~nm, assuming a sharp electron probe. Note that for this, the zero deflection of the measurement needs to be set to the deflection representing the uniformly magnetized background outside of the skyrmions. This is equivalent to the subtraction of the phase ramp described in the theoretical part. The phase's point of inflection $X_\Delta^\Phi$ is located at the minimum of the $\text{DPC}_\text{X}$ signal and gives a domain wall width $\Delta= 84 \pm 27$~nm. Using these two parameters, the skyrmion structure can be estimated using equations \ref{frml_skstructure1}-\ref{frml_skstructure4}.

To study the influence of electron-optical aberrations that lead to an increased probe size, further simulations have been conducted. For this, the center of mass from simulated ronchigrams is calculated with respect to probe radius $R_\text{probe}$ for each pixel of the N\'eel skyrmion's electron phase. The resulting images look very similar to the ones shown in Fig.~\ref{figure3}(a), although key features like the zero, or minimum phase gradient are subject so slight shifts. When measuring with large probesizes, these shifts will lead to erratic skyrmion sizes when applying the proposed rules in the previous sections. To compensate for this, correction terms for $R_\text{exp}$ and $(\Delta/R)_\text{exp}$ are calculated from the simulations, which are shown in Fig.~\ref{figure3}(d). In the presented experiment a large aperture with a diameter of $100~\mu\text{m}$ led to a rather low resolution of around $25~\text{nm}$. This gives a probe radius that is around half of the skyrmion radius. Therefore, the actual skyrmion radius is expected to be around 13\% smaller and $\Delta / R$ decreases by around $0.1$.

\section{Summary}

In this work, the spin structure of isolated magnetic skyrmions is related to the electron phase modulation obtained from Lorentz microscopy. The two most common skyrmion types, the Bloch and N\'eel skyrmion, are addressed. The universal results are independent of saturation magnetization, skyrmion size and typical TEM sample thicknesses. For each of the two skyrmion types, convenient rules are presented to determine the skyrmion radius and wall width that define the skyrmion configuration in the used model without the need of further simulations. These findings might provide useful insights for material scientists developing tailored skyrmion materials with engineered skyrmion configurations. Additionally, first DPC measurements of room temperature N\'eel skyrmion samples that match the theoretical expectations are presented. The measurements showcase an excellent application for the theoretical part of this work. The proposed method for quick skyrmion structure characterization provides a groundwork for many additional techniques that require the actual skyrmion's structure like Ferromagnetic Resonance \cite{Schwarze2015, Poellath2019} or Skyrmion-Hall-Effect measurements \cite{Sampaio2013}.

\textit{Acknowledgements} We wish to thank K. Fallon for fruitful discussions. S.P.\ C.B.\ and J.Z.\ acknowledge funding by the Deutsche Forschungsgemeinschaft  (DFG, German Research Foundation) via SPP2137. 
This project has received funding from the European Metrology Programme for Innovation and Research (EMPIR) co-financed by the Participating States and from the European Union’s Horizon 2020 research and innovation programme. This  work  has  also been  funded  by  the  Deutsche Forschungsgemeinschaft (DFG, German Research Foundation) under Germany’s Excellence Strategy EXC-2111 390814868. T.L.\ , N.L.\ and W. Z.\ acknowledge funding by the National Natural Science Foundation of China (Grants No.\ 11574018, and No. 61627813), the International Collaboration
Project (Grant No.\ B16001), the National Key
Technology Program of China (Grant No.\ 2017ZX01032101).


%

\end{document}